\documentclass[twocolumn,twoside,slac_two]{revtex4}
\usepackage{graphicx}
\usepackage{fancyhdr}
\begin{document}

\title{Design and Tests of the Hard X-ray Polarimeter X-Calibur}

%

\author{M.~Beilicke$^{*}$, W.R.~Binns, J.~Buckley, R.~Cowsik,
P.~Dowkontt, A.~Garson, Q.~Guo, M.H.~Israel, K.~Lee, H.~Krawczynski}
\affiliation{Department of Physics and McDonnell Center for the Space
Sciences, Washington University in St.~Louis, MO, USA ($^{*}$E-mail:
beilicke@physics.wustl.edu)}

\author{M.G.~Baring}
\affiliation{Rice University, TX, USA}

\author{S.~Barthelmy, T.~Okajima, J.~Schnittman, J.~Tueller}
\affiliation{Goddard Space Flight Center, MD, USA}

\author{Y.~Haba, H.~Kunieda, H.~Matsumoto, T.~Miyazawa, K.~Tamura}
\affiliation{Nagoya University, Japan}

\begin{abstract}

X-ray polarimetry promises to give new information about high-energy
astrophysical sources, such as binary black hole systems, micro-quasars,
active galactic nuclei, and gamma-ray bursts. We designed, built and
tested a hard X-ray polarimeter {\it X-Calibur} to be used in the focal
plane of the InFOC$\mu$S grazing incidence hard X-ray telescope.
X-Calibur combines a low-Z Compton scatterer with a CZT detector
assembly to measure the polarization of $10-80 \, \rm{keV}$ X-rays
making use of the fact that polarized photons Compton scatter
preferentially perpendicular to the electric field orientation.
X-Calibur achieves a high detection efficiency of order unity.
\end{abstract}

\maketitle

\thispagestyle{fancy}


\section{INTRODUCTION}
\label{sec:intro}

{\bf Motivation.} Spectro-polarimetric X-ray observations are capable of
providing important information to study the non-thermal emission
processes of various astrophysical sources~-- namely the fraction and
orientation of linearly polarized photons \cite{Krawcz2011}. So far,
only a few missions have successfully measured polarization in the soft
(OSO-8 \cite{Weisskopf1978}) and hard (Integral \cite{Dean2008}) X-ray
energy regime. The Crab nebula is the only source for which the
polarization of the X-ray emission has been established with a high
level of confidence \cite{Weisskopf1978, Dean2008}. Integral
observations of the X-ray binary Cygnus\,X-1 indicate a high fraction of
polarization in the hard X-ray/gamma-ray bands \cite{Laurent2011}.

{\bf Future missions.} There are currently no missions in orbit that are
capable of making sensitive X-ray polarimetric observations. This will
change by the launch of the satellite-borne {\it Gravity and Extreme
Magnetism SMEX} (GEMS) mission \cite{GEMS} scheduled for 2014 ($2 - 10
\, \rm{keV}$).  The {\it Soft Gamma-Ray Imager} on {\it ASTRO-H}
\cite{Tajima2010} will also have capabilities of measuring polarization,
but the results may be plagued by systematic uncertainties. The hard
X-ray polarimeter {\it X-Calibur} discussed in this paper has the
potential to cover the energy range above $10 \, \rm{keV}$, combining a
high detection efficiency with a low level of background and
well-controlled systematic errors.

{\bf Scientific potential.} Synchrotron emission results in linearly
polarized photons (electric field oriented perpendicular to the magnetic
field lines). The polarized synchrotron photons can be inverse-Compton
(IC) scattered by relativistic electrons~-- weakening the fraction of
polarization and imprinting a scattering angle dependence
\cite{Poutanen1994} to the observed fraction of polarization. Other
important mechanism for polarizing photons are Thomson scattering and
curvature radiation. The scientific potentials of spectro-polarimetric
hard X-ray observations are (see Krawczynski et al. (2011) and
references therein \cite{Krawcz2011}):

{\it 1) Binary black hole (BH) systems.} Relativistic aberration and
beaming, gravitational lensing, and gravitomagnetic frame-dragging will
result in an energy-dependent fraction of X-ray polarization from a
Newtonian accretion disk \cite{Connors1977} since photons with higher
energies originate closer to the BH than the lower-energy photons. 
Schnittman and Krolik \cite{Schnittman2009, Schnittman2010} calculate
the expected polarization signature including the effects of deflection
of photons emitted in the disk by the strong gravitational forces in the
regions surrounding the black hole and of the re-scattering of these
photons by the accretion disk. Spectro-polarimetric observations can
constrain the mass and spin of the BH \cite{Schnittman2009}, as well as
the inclination of the inner accretion disk and the shape of the corona
\cite{Schnittman2010}.

{\it 2) Pulsars and pulsar wind nebulae.} High-energy particles in
pulsar magnetospheres are expected to emit synchrotron and/or curvature
radiation which are difficult to distinguish from one another. However,
since the orbital planes for accelerating charges that govern these two
radiation processes are orthogonal to each other, their polarized
emission will exhibit different behavior in position angle and
polarization fraction as functions of energy and the phase of the pulsar
\cite{Dean2008}. In magnetars, the highly-magnetized cousins of pulsars,
polarization-dependent resonant Compton up-scattering is a leading
candidate for generating the observed hard X-ray tails. In both these
classes of neutron stars, phase-dependent spectro-polarimetry can probe
the emission mechanism, and provide insights into the magnetospheric
locale of the emission region.

{\it 3) Relativistic jets in active galactic nuclei (AGN).} Relativistic
electrons in jets of AGN emit polarized synchrotron radiation at
radio/optical wavelengths. The same electron population is believed to
produce hard X-rays by inverse-Compton scattering of a photon field.
Simultaneous measurements of the polarization angle and the fraction of
polarization in the radio to hard X-ray band could help to study (i)
synchrotron self-Compton models (X-ray polarization tracks polarization
at radio/optical wavelengths in fraction and orientation angle)
\cite{Poutanen1994} vs. (ii) external-Compton models for which the hard
X-rays will have a relatively small ($<$10\%) fraction of polarization
\cite{McNamara2009}.  Polarization also allows one to test the structure
(i.e. helical) of the magnetic field of the jet.

{\it 4) Gamma-ray bursts (GRBs)}. GRBs are believed to be connected to
hyper-nova explosions and the formation of relativistic jets. As in AGN
jets, the structure and particle distribution responsible for GRBs can
be revealed by X-ray polarization measurements.
 
{\it 5) Lorentz invariance.} X-ray polarimetric observations can be used
to test theories violating Lorentz invariance \cite{Fan2007} with
unprecedented accuracies by probing the helicity dependence of the speed
of light.

\section{DESIGN OF X-CALIBUR}
\label{sec:design}

\begin{figure}[t]
   \includegraphics[width=0.55\textwidth, angle=90]{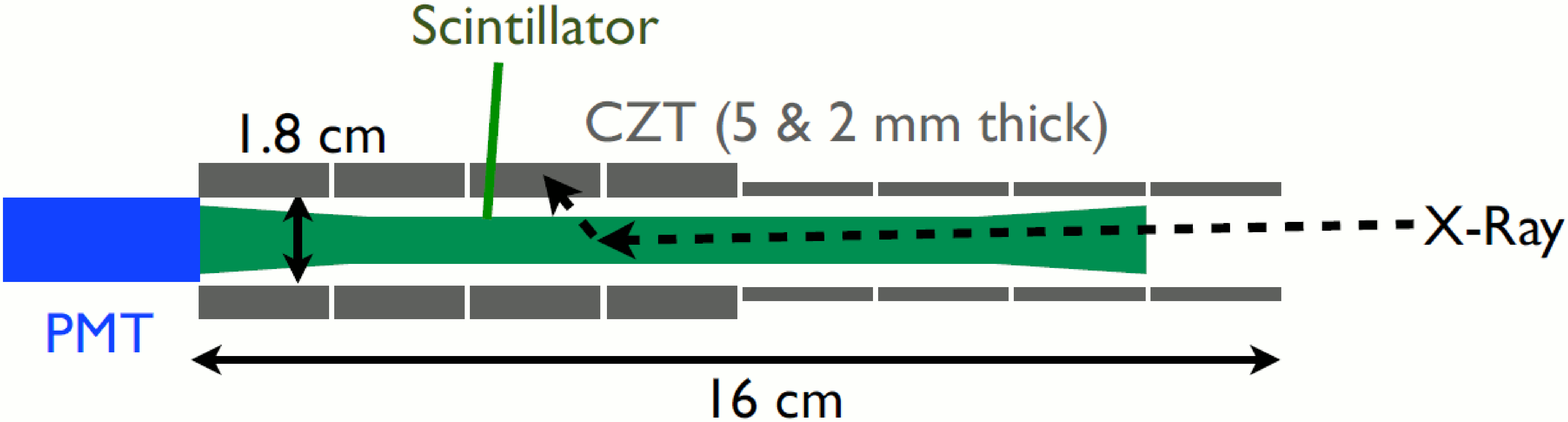}
   \hfill
   \includegraphics[width=0.30\textwidth]{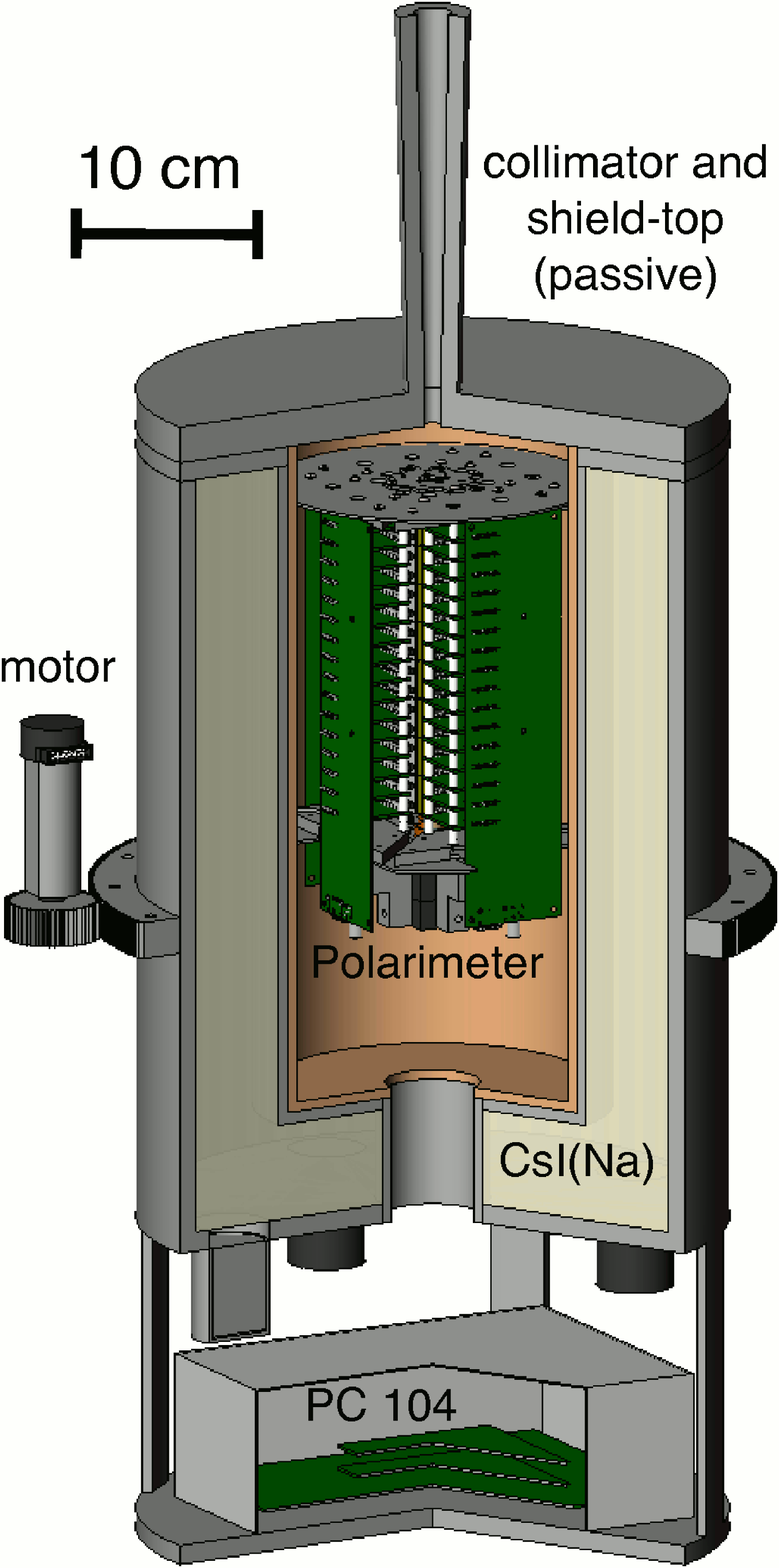}

\caption{\label{fig:Design} {\bf Left:} Conceptual design of X-Calibur:
Incoming X-rays are Compton-scattered in the scintillator rod (read by a
PMT) and are subsequently photo-absorbed in one of the CZT detectors
surrounding the rod. {\bf Right:} X-Calibur design with read-out
electronics, shielding and azimuthal rotation bearing.}

\end{figure}

Figure~\ref{fig:Design} illustrates the conceptual design of the
X-Calibur polarimeter. A low-Z scintillator is used as
Compton-scatterer. For sufficiently energetic photons, the Compton
interaction produces a measurable scintillator signal which is read by a
PMT. The scattered X-rays are photo-absorbed in surrounding rings of
high-Z Cadmium Zinc Telluride (CZT) detectors. This combination of
scatterer/absorber leads to a high fraction of unambiguously detected
Compton events. Linearly polarized X-rays will preferably
Compton-scatter perpendicular to their E field vector~-- resulting in a
modulation of the azimuthal event distribution.

\begin{figure}
   \includegraphics[width=0.28\textwidth]{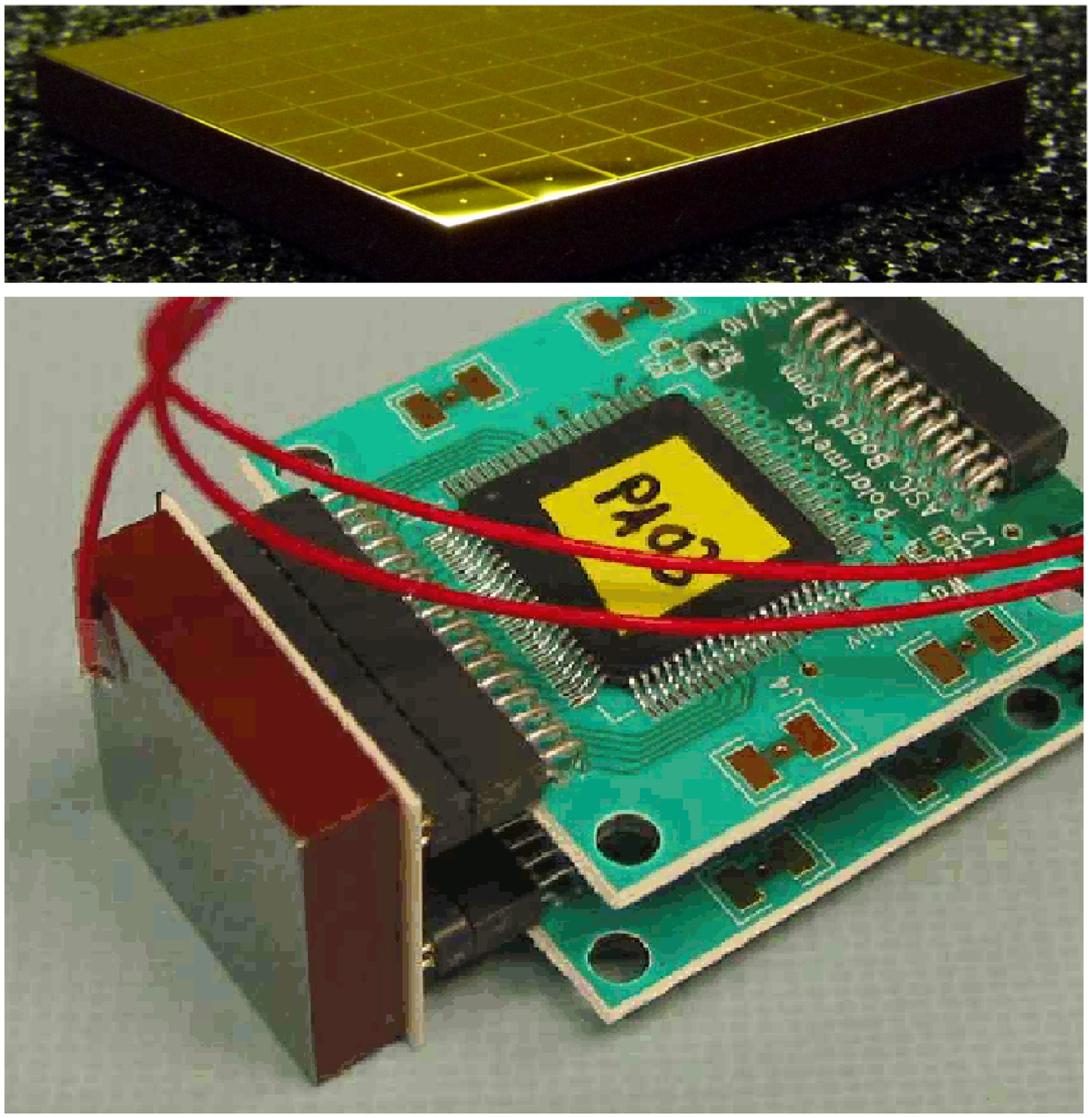}
   \hfill
   \includegraphics[width=0.19\textwidth]{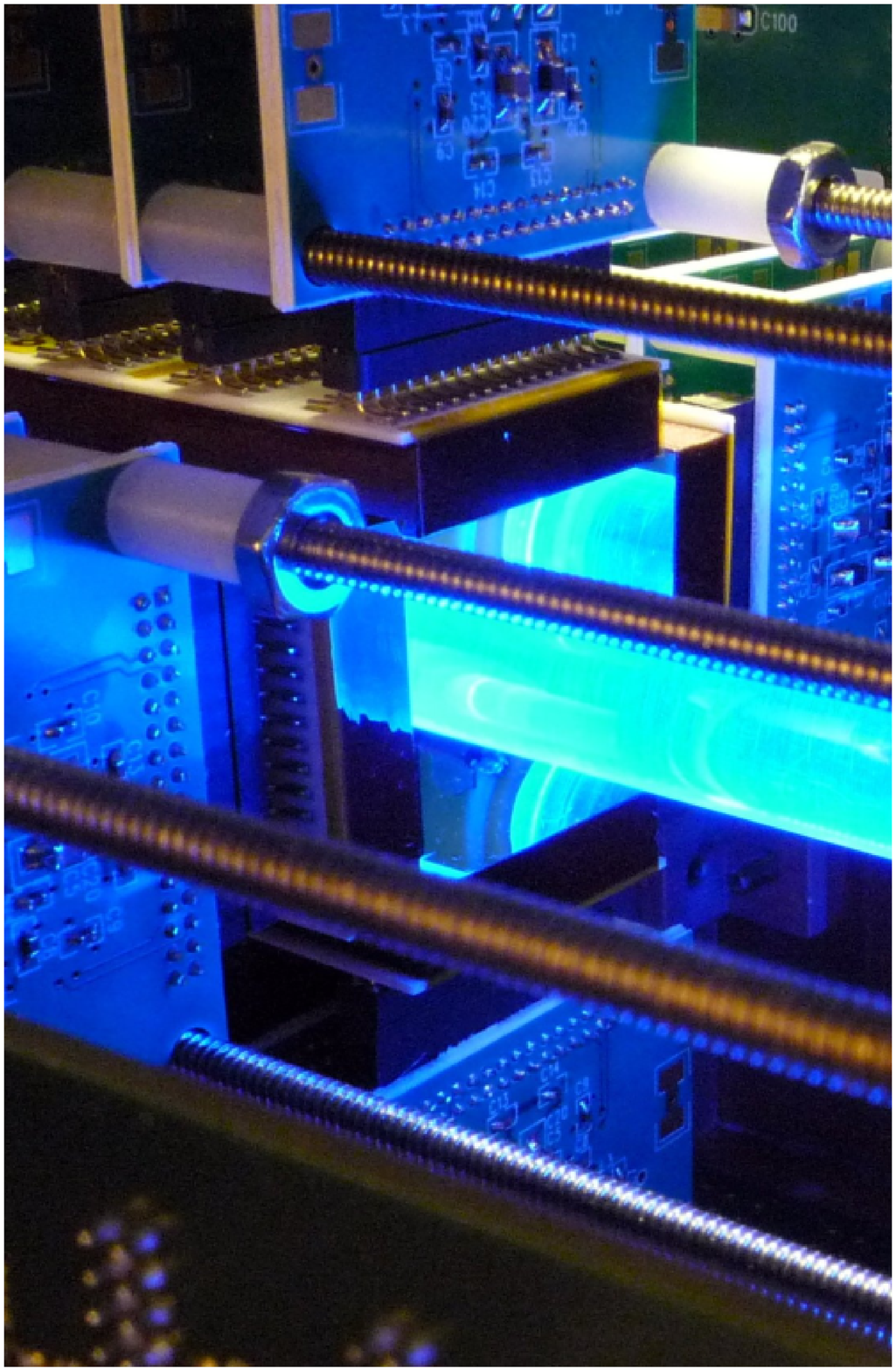}

\caption{\label{fig:Fotos} {\bf Left:} Top: $2 \times 2 \times 0.2 \,
\rm{cm}^{3}$ CZT detector (64 anode pixels). Bottom: $2 \times 2 \times
0.5 \, \rm{cm}^{3}$ detector bonded to a ceramic chip carrier, plugged
into 2 ASIC readout boards. The high-voltage cable is glued to the
detector cathode (red wire). {\bf Right:} Laboratory version of
X-Calibur. The scintillator (blueish glow) is surrounded by two detector
rings~-- each consisting of four 64 pixel CZT detectors.}

\end{figure}

Each CZT detector ($2 \times 2 \, \rm{cm}^{2}$) is contacted with a
64-pixel anode grid and a monolithic cathode facing the scintillator
rod. Two detector thicknesses ($0.2 \, \rm{cm}$ and $0.5 \, \rm{cm}$)
are being tested in the current setup. Each CZT detector is permanently
bonded (anode side) to a ceramic chip carrier which is plugged into the
readout board. Figure~\ref{fig:Fotos} (left) shows a single CZT detector
unit as well as the readout electronics. Each CZT detector is read out
by two digitizer boards (32 channel ASIC developed by G.~De~Geronimo
(BNL) and E.~Wulf (NRL) \cite{Wulf2007} and a 12-bit ADC). 16 digitizer
boards (8 CZT detectors) are read out by one harvester board
transmitting the data to a PC-104 computer. Four CZT detector units form
a ring surrounding the scintillator slab. The scintillator EJ-200 is
read by a Hamamatsu R7600U-200 PMT. The PMT trigger information allows
to effectively select scintillator/CZT events from the data which
represent likely Compton-scattering candidates. The polarimeter and the
front-end readout electronics will be located inside an active CsI(Na)
anti-coincidence shield with $5 \, \rm{cm}$ thickness and a passive lead
shield/collimator at the top (Fig.~\ref{fig:Design}) to suppress charged
and neutral particle backgrounds.

We plan to use the X-Calibur polarimeter in the focal plane of the
InFOC$\mu$S experiment \cite{InFocus_FirstFlight}. A Wolter grazing
incidence mirror focuses the X-rays on the X-Calibur polarimeter. In
order to reduce the systematic uncertainties (including biases generated
by the active shield, a possible pitch of the polarimeter with respect
to the X-ray telescope, etc.), the polarimeter and the active shield
will be rotated around the optical axis using a ring bearing (see
Fig.~\ref{fig:Design}). Counter-rotating masses will be used to cancel
the net-angular momentum of the polarimeter assembly during the flight.
The advantages of the X-Calibur/InFOC$\mu$S design are (i) a high
detection efficiency by using more than $80 \%$ of photons impinging on
the polarimeter, (ii) low background due to the usage of a focusing
optics instead of large detector volumes, and (iii) minimization and
control of systematic effects.

\section{SIMULATIONS}
\label{sec:simulations}

Simulations of the X-Calibur polarimeter were performed using the {\it
Geant4} package \cite{GEANT} with the Livermore low-energy
electromagnetic model list. A balloon flight in the focal plane of the
InFOC$\mu$S mirror assembly was assumed. We accounted for atmospheric
absorption at a floating altitude of $130,000$ feet using the NIST XCOM
attenuation coefficients \cite{NIST_XCOM} and an atmospheric depth of
$2.9 \, \rm{g/cm}^{2}$ (observations performed at zenith). A Crab-like
source was simulated.  The X-Calibur modulation factor is $\mu = 0.52$
for a $100 \%$ polarized beam. The minimum detectable polarization
\cite{Krawcz2011} in the $10-80 \, \rm{keV}$ energy range will be $4 \%$
assuming $5.6 \, \rm{hr}$ of on-source observations of a Crab-like
source combined with a 1.4~hr background observation of an adjacent
empty field. More details about the simulations can be found in Guo et
al. (2010) \cite{Guo2010}.

\begin{figure*}
 \includegraphics[width=0.50\textwidth]{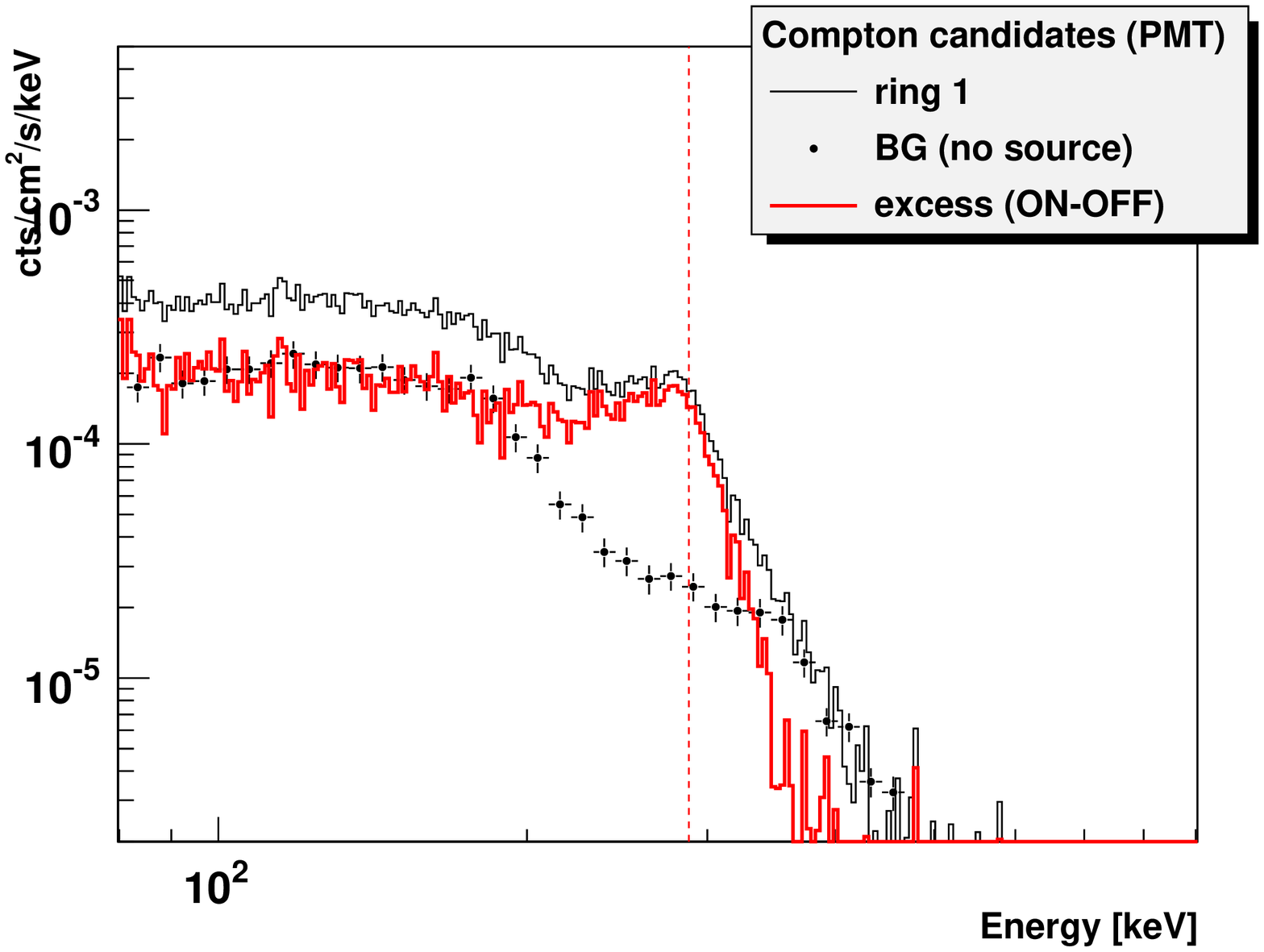}
 \hfill
 \includegraphics[width=0.46\textwidth]{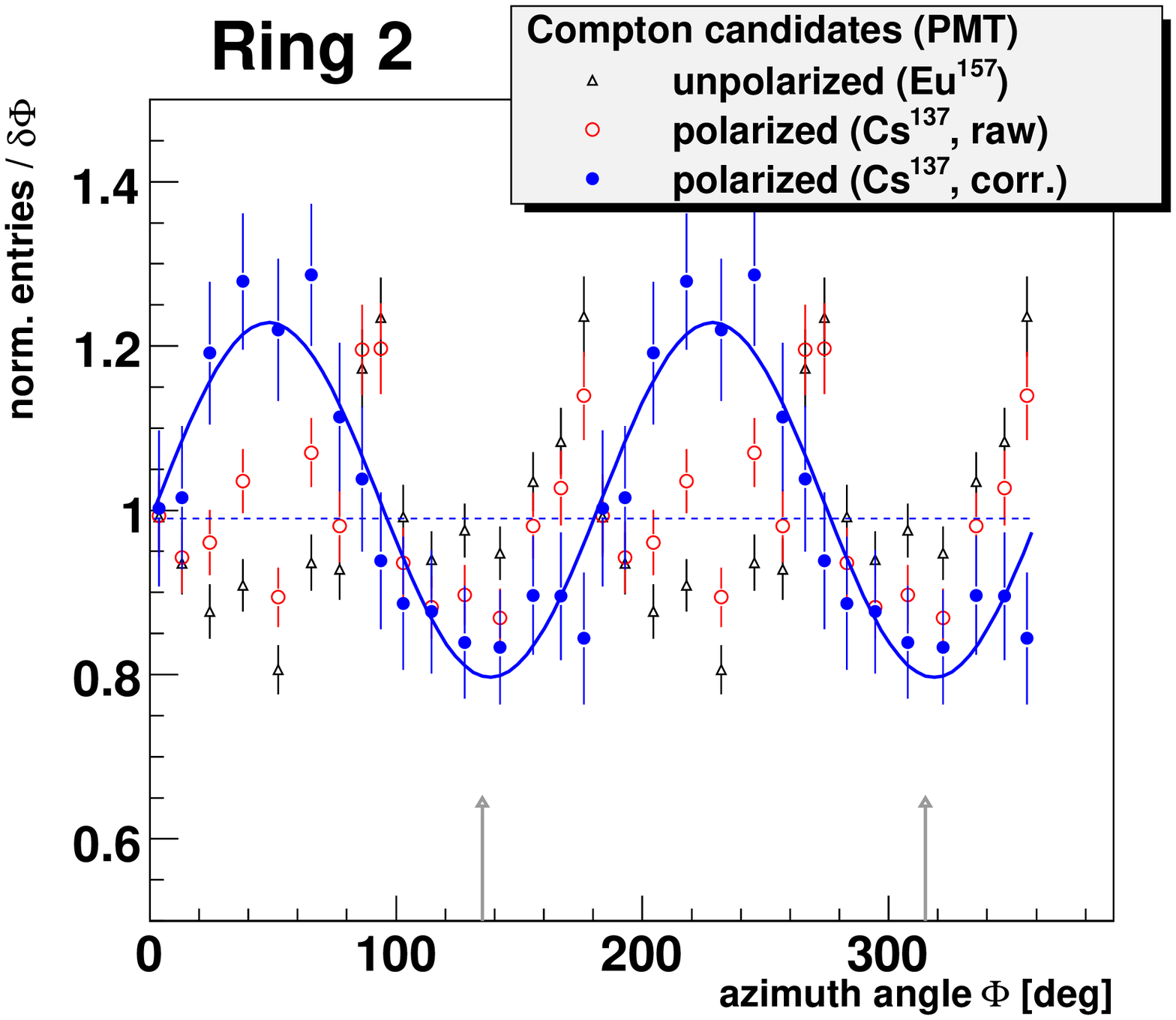}

\caption{\label{fig:Spectra} {\bf Left:} Energy spectrum of $288 \,
\rm{keV}$ X-rays (partially polarized) after being scattered in the
scintillator (different scattering angles and corresponding energy
transfer). The background measurement is done without a source (cosmic
rays). The vertical line ($288 \, \rm{keV}$) indicates the energy of the
incoming X-ray beam. {\bf Right:} Azimuthal event distribution for the
second installed CZT detector ring. Shown are raw events of the
polarized beam (red), an unpolarized beam (black), and the acceptance
corrected polarized events (blue). A sine function ($180 \deg$
modulation) was fit to the corrected data of the polarized beam. The
vertical arrows indicate the plane of the electric field vector. The $90
\deg$ modulation of the unpolarized beam is a result of the 4-fold
geometry of the CZT detector assembly.}

\end{figure*}

\section{FIRST MEASUREMENTS}
\label{sec:FirstMeasurements}

Using funding from Washington University's McDonnell Center for the
Space Sciences, a flight-ready version of the X-Calibur polarimeter was
assembled and tested in the laboratory. First measurements were
performed with 3 detector rings installed ($0.5 \, \rm{cm}$ thickness).
Before installation, IV-curves were taken for all detector pixels,
followed by a calibration run using a Eu$^{152}$ source (lines at
$39.9$, $45.7$, $121.8$ and $344.3 \, \rm{keV}$). After calibration, a
collimated Eu$^{152}$ source was used to determine the azimuthal
X-Calibur acceptance for an unpolarized beam. Another data run was taken
without a source to determine the background induced by cosmic rays
secondaries hitting the detector assembly. Only CZT events with a
simultaneous ($30 \mu \rm{s}$) scintillator trigger are used.

A polarized beam was generated by scattering a strong Cs$^{137}$ source
(line at $662 \, \rm{keV}$) off a lead brick. A lead collimator allowed
only X-rays with a scattering angle of $\sim 90 \deg$ to enter the
polarimeter. The X-ray beam of the scattered photons has a mean energy
of $288 \, \rm{keV}$ and was polarized to $\sim 55 \%$ (modulation
factor of $\mu = 0.4$). The expected relative amplitude in the
normalized $\Phi$ distribution is $0.55 \times 0.4 = 0.22$.

Figure~\ref{fig:Spectra} (left) shows the raw spectrum of the first CZT
polarimeter ring measured from (i) the polarized beam, (ii) a background
spectrum measured without a source and (iii) the excess spectrum
corresponding to the energy spectrum of the scattered/polarized beam. As
expected, the excess spectrum drops off for energies higher than $288 \,
\rm{keV}$ (vertical line)~-- the energy of the $90 \deg$-scattered
Cs$^{137}$ photons entering the polarimeter. The continuum below this
energy is the result of $288 \, \rm{keV}$ photons being
Compton-scattered at different depths in the scintillator rod and
therefore being reflected to the first CZT ring under different
scattering angles and corresponding different Compton energy losses. The
little bump in the spectrum around $288 \, \rm{keV}$ may originate from
direct CZT hits without a Compton-scattering in the scintillator.

Figure~\ref{fig:Spectra} (right) shows the azimuthal scattering
distribution of the polarized and unpolarized beam for the second
installed CZT detector ring.  Only events with a simultaneous
scintillator trigger and with a deposited CZT energy between $100-330 \,
\rm{keV}$ are used (see spectrum in Fig.~\ref{fig:Spectra}, left). The
data of the polarized beam are corrected for the acceptance of the
polarimeter (derived from the unpolarized X-ray beam).  As expected for
a polarized beam, a $180 \deg$ modulation can be seen with a maximum of
azimuthal scattering angle perpendicular ($\Phi + 90 \deg$) to the plane
of the $E$ field vector of the polarized beam (indicated by the gray
arrows). A sine function was fit to the $\Phi$-distribution of the
polarized beam resulting in a relative amplitude of $0.22$. The data are
in excellent agreement with expectations.

\section{SUMMARY AND CONCLUSION}
\label{sec:summary}

We designed a hard X-ray polarimeter X-Calibur and studied its projected
performance and sensitivity for a 1-day balloon flight with the
InFOC$\mu$S X-ray telescope. X-Calibur combines a detection efficiency
of close to $100\%$ with a high modulation factor of $\mu \approx 0.5$,
as well as a good control over systematic effects.  X-Calibur was
successfully tested/calibrated in the laboratory with a polarized beam
of $288 \, \rm{keV}$ photons. We applied for a 1-day
X-Calibur/InFOC$\mu$S balloon flight. Our tentative observation program
includes galactic sources (Crab, Her\,X-1, Cyg\,X-1, GRS\,1915,
EXO\,0331) and one extragalactic source (Mrk\,421) for which sensitive
polarization measurements would be carried through. We envision
follow-up longer duration balloon flights (from the northern and
southern hemisphere), possibly using a mirror with increased area. In
the ideal case these flights would be performed while the GEMS mission
is in orbit to achieve simultaneous coverage in the $0.5 - 80 \,
\rm{keV}$ regime. Successful balloon flights would motivate a
satellite-borne hard X-ray polarimetry mission.

\bigskip 
\begin{acknowledgments}

We are grateful for NASA funding from grant NNX10AJ56G and discretionary
founding from the McDonnell Center for the Space Sciences to build the
X-Calibur polarimeter. Q.Guo thanks the Chinese Scholarship Council for
the financial support (NO.2009629064, stay at Washington University).

\end{acknowledgments}


\end{document}